\documentclass{article}

\usepackage{arxiv}

\usepackage{hyperref}       % hyperlinks
\usepackage{xcolor}
\hypersetup{colorlinks=true, linkcolor=blue, citecolor=blue}
\usepackage{booktabs}       % professional-quality tables
\usepackage{placeins}       % \FloatBarrier: keeps the floats of Section 3.3 before Section 3.4
\usepackage{amsfonts}       % blackboard math symbols
\usepackage{amsmath}
\usepackage{amssymb}
\usepackage{nicefrac}       % compact symbols for 1/2, etc.
\usepackage{microtype}      % microtypography
\usepackage{ifpdf}
  \usepackage{graphicx}
\usepackage[numbers,sort&compress]{natbib}   % numbered citations with \cite{}
\usepackage{doi}
\usepackage{siunitx}
\usepackage{mathcomp}
\newcommand{\bm}[1]{{\mbox{\boldmath $#1$}}}
\newcommand{\GP}{\mathrm{GP}}
\newcommand{\RP}{\mathrm{RP}}
\newcommand{\OP}{\mathrm{OP}}
\newcommand{\Rey}{\mathrm{Re}}
\newcommand{\pl}[1]{(\textit{#1})}
\newcommand{\panel}[2]{\begin{tabular}[b]{@{}l@{}}\pl{#1}\\[1pt]#2\end{tabular}}

\title{A branch-free three-dimensional\\ ghost-cell immersed boundary method with\\ direction-averaged extrapolation kernels}

\newif\ifuniqueAffiliation
\uniqueAffiliationtrue

\ifuniqueAffiliation % Standard variant of author block
\author{Itsuki Takarabe\\
        Department of Mechanical Engineering,
        Keio University\\
        itsuki.takarabe@keio.jp\\
\And
 Ryuto Saeki \\
 	Department of Mechanical Engineering,
	Keio University\\
	saekiryuto@keio.jp\\
\And
 Koji Fukagata \\
	Department of Mechanical Engineering,
	Keio University\\
	fukagata@mech.keio.ac.jp\\
}

\else
\usepackage{authblk}

\newbox{\orcid}\sbox{\orcid}{\includegraphics[scale=0.06]{orcid.pdf}}
\author[1]{%
	\href{https://orcid.org/0000-0000-0000-0000}{\usebox{\orcid}\hspace{1mm}David S.~Hippocampus\thanks{\texttt{hippo@cs.cranberry-lemon.edu}}}%
}
\author[1,2]{%
	\href{https://orcid.org/0000-0000-0000-0000}{\usebox{\orcid}\hspace{1mm}Elias D.~Striatum\thanks{\texttt{stariate@ee.mount-sheikh.edu}}}%
}
\affil[1]{Department of Computer Science, Cranberry-Lemon University, Pittsburgh, PA 15213}
\affil[2]{Department of Electrical Engineering, Mount-Sheikh University, Santa Narimana, Levand}
\fi

\renewcommand{\headeright}{}
\renewcommand{\undertitle}{}
\makeatletter
\renewcommand{\shorttitle}{{\def\\{\unskip\space\ignorespaces}\@title}}
\makeatother

\usepackage{etoolbox}
\makeatletter
\patchcmd{\@maketitle}{\vskip 0.4in \@minus 0.1in}{\vskip 0.25in}{}{}
\patchcmd{\@maketitle}{\vskip 0.2in}{\vskip 0.15in}{}{}
\patchcmd{\@maketitle}{\rule{\z@}{24\p@}}{\rule{\z@}{18\p@}}{}{}
\patchcmd{\@maketitle}{\rule{\z@}{24\p@}}{\rule{\z@}{18\p@}}{}{}
\patchcmd{\@maketitle}{\rule{\z@}{24\p@}}{\rule{\z@}{18\p@}}{}{}
\makeatother

\begin{document}
\maketitle
\baselineskip 14pt

\begin{abstract}
Ghost-cell immersed boundary methods impose the wall boundary condition on a Cartesian grid by assigning values to the grid points just inside the body, the ghost points.
A common approach reconstructs the ghost value from an image point, which requires a wall normal for every ghost point.
Special treatments are also often required where the interpolation stencil leaves the fluid region, e.g., near thin, concave or highly curved parts of complex bodies.
We present a ghost-cell method that needs neither wall normals nor image points.
Instead of a single extrapolation direction, the method uses all fluid nodes within 2.5 grid spacings of a ghost point that satisfy a simple geometric condition.
The ghost value is obtained as the average of one-dimensional linear extrapolations from these nodes through the wall crossings of the connecting segments.
The weights depend only on the geometry and are computed offline as one kernel on a $5 \times 5 \times 5$ stencil per ghost point.
The boundary treatment at run time is thus an explicit, branch-free weighted sum.
For the flows around a sphere and the Stanford Bunny, a self-convergence study shows approximately second-order convergence of the velocity in the $L_1$ and $L_2$ norms.
The maximum difference also decreases at second order for the sphere but more slowly for the Stanford Bunny, where it is observed at a few nodes next to the wall.
The drag coefficient of the sphere at Reynolds numbers of 25--200 agrees with a previous study within 2.4~\% for 32 grid cells per diameter and 0.4~\% for 64.
For the Stanford Bunny at a Reynolds number of 100, the force coefficients agree with a body-fitted solution within 1~\% for 64 grid cells per reference length, without any geometry-specific treatment.
Although no pressure condition is imposed on the wall, the pressure field is smooth and consistent with the body-fitted solution.
\end{abstract}
% keywords can be removed
\keywords{Direct numerical simulation \and Immersed boundary method \and Ghost-cell method \and Complex geometry }

\baselineskip 16pt

\section{Introduction}
\label{sec:intro}

Flows around bodies of complex shape are commonly simulated on body-fitted or unstructured grids.
For three-dimensional geometries, however, the generation of such grids is a major part of the total effort.
Immersed boundary methods (IBMs) avoid this step \cite{peskin1972,mittal2005,verzicco2023}.
They discretize the governing equations on a Cartesian grid that does not conform to the body, and they impose the wall boundary condition by modifying the equations or the discrete solution near the wall.
As a result, the grid generation becomes trivial, and fast structured solvers remain available, e.g., direct Poisson solvers based on the fast Fourier transform.

The existing boundary treatments differ in how they represent the wall.
In the continuous forcing approach \cite{peskin1972,goldstein1993,peskin2002} and in the volume penalization method \cite{angot1999}, a body force is distributed over the grid points around or inside the body.
These methods are simple, but they do not impose the boundary condition exactly on the wall.
The continuous forcing smears the wall over a few grid cells, so that the accuracy near the wall is of first order \cite{peskin2002}.
The penalization recovers the boundary condition only in the limit of a vanishing penalization parameter \cite{angot1999}.
In the discrete (direct) forcing approach \cite{mohdyusof1997,fadlun2000,kim2001,balaras2004}, on the other hand, the velocity at the grid points next to the wall is chosen so that the boundary condition is met on the wall.
This keeps the interface sharp and the accuracy of second order.
The reconstruction, however, needs an interpolation direction or stencil to be defined for every forcing point, and the treatments are often not unified on a three-dimensional surface.
Ghost-cell methods \cite{tseng2003,mittal2008} are a variant of this approach.
They assign values to the grid points just inside the body, the ghost points, so that the discrete operators of the flow solver can be used without modification at all fluid points.
Cut-cell methods \cite{hartmann2011} take yet another route.
They reshape the cells that are cut by the wall and satisfy the conservation laws in them.
The drawback is a complicated cell geometry in three dimensions.

In the widely used ghost-cell method of Mittal et al.~\cite{mittal2008}, each ghost point is mirrored across the wall along the wall normal.
The value at the resulting image point is interpolated tri-linearly from the eight surrounding grid points in three-dimensional cases, and the ghost value then follows from the boundary condition at the midpoint, which lies on the wall.
On complex three-dimensional shapes, however, this procedure meets three difficulties.
First, the point on the wall has to be found for every ghost point, but it is not always unique, and a normal intercept may not exist.
Mittal et al.~\cite{mittal2008} therefore search for it with several fall-back steps.
Second, one of the eight grid points is often the ghost point itself, and its equation is then replaced by the boundary condition as a special case.
Third, other ghost points may also be among the eight points, so that the ghost values are coupled and have to be solved together.
Both cases occur, e.g., at a concave corner (Fig.~\ref{fig:corner}\pl{a}).
A related approach is that of Tseng and Ferziger~\cite{tseng2003}, which extrapolates the ghost value through the wall from fluid points.
They reported that this extrapolation becomes unstable when a fluid point of the stencil lies close to the wall, and they proposed solutions such as the use of an image point.

One way to avoid the wall normal is to apply the boundary condition direction by direction.
The local directional ghost-cell approach of Berthelsen and Faltinsen~\cite{berthelsen2008} and the directional ghost-cell method of Chi et al.~\cite{chi2020} follow this idea.
They extrapolate the ghost value along each grid line separately, using the point where that grid line crosses the wall.
This one-dimensional construction is simple and robust.
However, a ghost point then has a different value for each direction, e.g., three values in three dimensions.
The discretization next to the wall therefore has to be changed, so that each difference uses its own ghost value \cite{berthelsen2008} or the result enters as a discrete forcing term \cite{chi2020}.

In another approach, Kor et al.~\cite{kor2017} proposed what they call the unified interpolation stencil.
It is a fixed sequence of one-dimensional interpolations along the wall normal, so that no case-by-case selection of the interpolation points is needed.
It was later extended to moving boundaries \cite{kor2018}.
These studies considered two-dimensional flows.
Table~\ref{tab:methods} summarizes how these methods obtain the ghost value.

\begin{table}[t]
\caption{Reconstruction of the ghost value in sharp-interface methods related to the present one.}
\label{tab:methods}
\centering
\small
\begin{tabular}{@{}p{0.23\textwidth}p{0.26\textwidth}p{0.20\textwidth}p{0.23\textwidth}@{}}
\toprule
Method & Ghost value obtained from & Geometric data per ghost point & Coupling and special cases \\
\midrule
Tseng and Ferziger~\cite{tseng2003} & extrapolation through the wall point; image point where needed & wall normal, image point & special treatment when the wall is close to a fluid point \\[2pt]
Mittal et al.~\cite{mittal2008} & image point with tri-linear interpolation (eight points) & wall normal, body intercept & ghost points in the stencil: coupled ghost values \\[2pt]
Berthelsen and Faltinsen~\cite{berthelsen2008}; Chi et al.~\cite{chi2020} & one-dimensional extrapolation along each grid line & wall crossings of the grid lines & direction-dependent (multi-valued) ghost points \\[2pt]
Kor et al.~\cite{kor2017} & fixed sequence of one-dimensional interpolations along the normal & wall normal, image point, intersection points & stencil must lie in the fluid (local grid refinement at highly concave walls) \\[2pt]
Present & average of one-dimensional linear extrapolations to the selected nearby fluid points & wall crossings of the connecting segments (offline) & none: explicit, branch-free weighted sum \\
\bottomrule
\end{tabular}
\end{table}

The objective of the present study is a ghost-cell method that remains simple on arbitrary three-dimensional geometries.
Specifically, the method should need no wall normal, no image point, no coupling between ghost points and no alternative treatment for particular shapes.
In addition, each ghost point should keep a single value, so that the flow solver remains unchanged near the wall.
The boundary treatment at run time should also be cheap and the same for every ghost point, whatever the local shape of the wall.
To meet these requirements, we average one-dimensional extrapolations over all nearby directions instead of searching for one good direction, and we compute the weights of this average before the simulation.
We then examine whether the method keeps second-order accuracy and agrees with previous studies and with a body-fitted solution, not only for a sphere but also for a body with thin and concave parts.

The rest of this paper is organized as follows.
Section~\ref{sec:method} describes the method.
Section~\ref{sec:results} then examines it for the flows around a sphere and around the Stanford Bunny.
There, we compare the drag coefficient of the sphere with previous studies and measure the order of convergence of the flow field for both geometries.
We then compare the flow around the Stanford Bunny with a body-fitted solution.
Section~\ref{sec:concl} summarizes the findings and the remaining limitations of the method.

\section{Numerical method}
\label{sec:method}

\subsection{Governing equations and flow solver}
\label{sec:solver}

We consider the incompressible flow of a Newtonian fluid, which is governed by the continuity and Navier--Stokes equations,
\begin{align}
\bm{\nabla} \cdot \bm{u} &= 0, \label{eq:cont}\\
\frac{\partial \bm{u}}{\partial t} &= - \bm{\nabla} \cdot (\bm{u}\bm{u}) - \bm{\nabla} p + \frac{1}{\Rey} \nabla^2 \bm{u}, \label{eq:ns}
\end{align}
where $\bm{u} = (u, v, w)$ is the velocity and $p$ is the pressure.
All quantities are made dimensionless with the free-stream velocity $U_\infty^*$, a reference length $D^*$ and the density $\rho^*$; for a sphere, $D^*$ is the diameter.
The Reynolds number is then $\Rey = U_\infty^* D^* / \nu^*$, where $\nu^*$ is the kinematic viscosity and the asterisk denotes dimensional quantities.

The equations are discretized on a staggered Cartesian grid with the energy-conserving second-order central difference scheme.
The flow solver is based on the code of Fukagata et al.~\cite{fukagata2006}.
For the time integration, the convective terms are treated with the low-storage third-order Runge--Kutta scheme and the viscous terms with the Crank--Nicolson scheme \cite{spalart1991,rai1991}.
The pressure is coupled by a fractional step method in the pressure-increment form of the SMAC method \cite{amsden1970}.
With $\bm{N}(\bm{u}) = -\bm{\nabla} \cdot (\bm{u}\bm{u})$, the substep $k$ $(= 1, 2, 3)$ of a time step $\Delta t$ reads
\begin{align}
\frac{\bm{u}^{\ast} - \bm{u}^{k-1}}{\Delta t} &= \gamma_k \bm{N}(\bm{u}^{k-1}) + \zeta_k \bm{N}(\bm{u}^{k-2}) + \alpha_k \left[ -\bm{\nabla} p^{k-1} + \frac{1}{2\Rey} \nabla^2 \left( \bm{u}^{\ast} + \bm{u}^{k-1} \right) \right], \label{eq:pred}\\
\nabla^2 \varphi &= \frac{\bm{\nabla} \cdot \bm{u}^{\ast}}{\alpha_k \Delta t}, \label{eq:poisson}\\
\bm{u}^{k} &= \bm{u}^{\ast} - \alpha_k \Delta t\, \bm{\nabla} \varphi, \qquad p^{k} = p^{k-1} + \varphi, \label{eq:corr}
\end{align}
with $\gamma_k = (8/15,\ 5/12,\ 3/4)$, $\zeta_k = (0,\ -17/60,\ -5/12$) and $\alpha_k = \gamma_k + \zeta_k$.
Equation~(\ref{eq:pred}) is solved for the increment $\bm{u}^{\ast} - \bm{u}^{k-1}$, and its implicit viscous terms are handled by approximate factorization into three one-dimensional tridiagonal systems \cite{kim1985}.
The Poisson equation~(\ref{eq:poisson}), on the other hand, is solved directly.
We use a cosine expansion in the streamwise direction $x$, a Fourier expansion in the spanwise direction $z$ and a tridiagonal solver in $y$.
For boundary conditions, a uniform velocity is prescribed at the inflow boundary, and a convective condition with a correction of the mass flux is used at the outflow boundary.
The boundaries in $y$ are free-slip walls, and the $z$ direction is periodic.
The grid is uniform in $x$ and $z$, as required by the direct Poisson solver.
In $y$, it has the same uniform spacing in a region around the body and is stretched outside when the domain is large.
In the following, $\Delta x$ denotes this grid spacing around the body.

\subsection{Ghost points and the one-dimensional extrapolation}
\label{sec:oned}

The body is described by a signed distance function $\psi(\bm{x})$, which is positive in the fluid region and negative in the solid region.
On the staggered grid, the three velocity components are defined at different locations.
The following construction is therefore carried out separately for each component, with $\psi$ evaluated at the nodes of that component; below, $u$ stands for any one of the components.
We define a ghost point (GP) as a solid node ($\psi \le 0$) that has at least one fluid node ($\psi > 0$) among its six face neighbors.

\begin{figure}[t]
\centering
\includegraphics[width=0.52\textwidth]{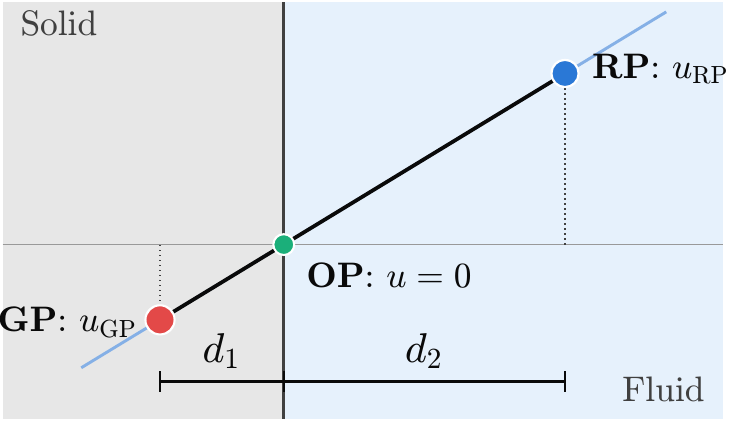}
\caption{One-dimensional building block: linear extrapolation from a reference point (RP) in the fluid to the ghost point (GP) through the object point (OP), where the segment between them crosses the wall.}
\label{fig:oned}
\end{figure}

The building block of the method is one-dimensional (Fig.~\ref{fig:oned}).
Consider a ghost point and one fluid node near it, which we call a reference point (RP).
The straight segment between them crosses the wall at the object point (OP).
Let $d_1$ be the distance from the GP to the OP, and $d_2$ the distance from the OP to the RP.
If the velocity varies linearly along the segment and satisfies the no-slip condition $u = 0$ at the OP, the value at the ghost point is
\begin{equation}
u_\GP = - \frac{d_1}{d_2}\, u_\RP . \label{eq:oned}
\end{equation}
For a wall that moves with the velocity $u_\OP$ at the object point, the same construction gives $u_\GP = [(d_1 + d_2)\, u_\OP - d_1\, u_\RP]/d_2$; in this paper, however, only stationary walls are considered.
The directional ghost-cell methods \cite{berthelsen2008,chi2020} use Eq.~(\ref{eq:oned}) along the grid lines.
Here, in contrast, it is applied along the segment to any fluid node, not only along the grid lines.

\subsection{Multi-directional averaging}
\label{sec:average}

\begin{figure}[t]
\centering
\begin{tabular}{@{}c@{\hspace{0.04\textwidth}}c@{}}
\panel{a}{\includegraphics[width=0.52\textwidth]{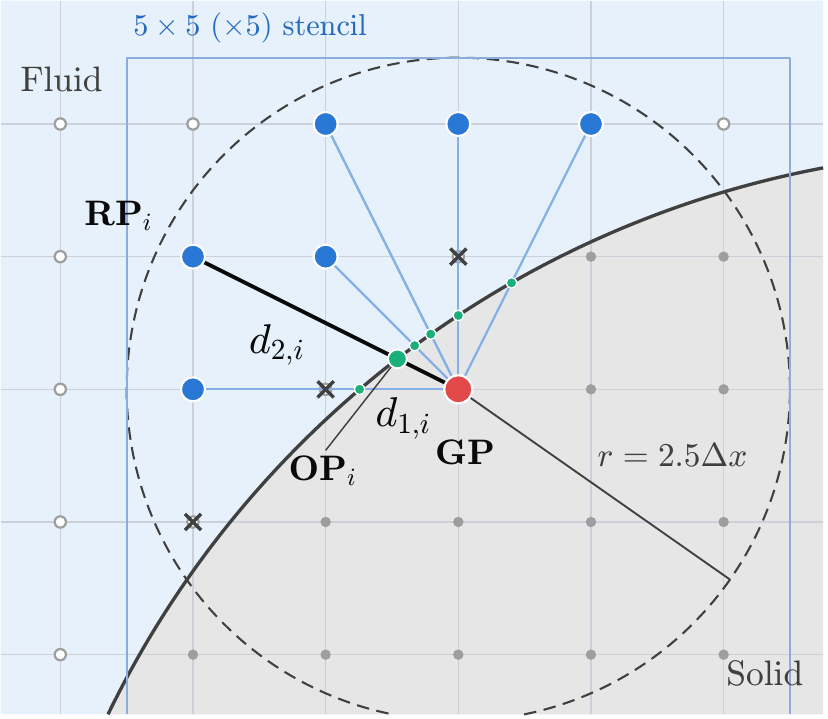}} &
\panel{b}{\includegraphics[width=0.34\textwidth]{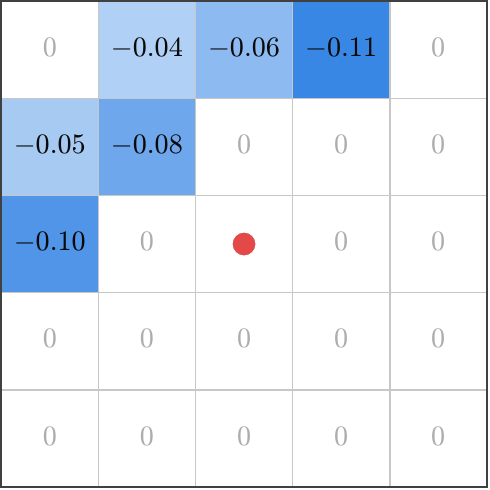}}
\end{tabular}
\caption{Multi-directional averaging, illustrated in two dimensions: \pl{a} a ghost point (GP), its reference points (RP$_i$, blue circles in the fluid within the radius $r$), the corresponding object points (OP$_i$, small circles on the wall) and the candidates rejected because $d_{1,i} \ge d_{2,i}$ (crosses); \pl{b} the resulting kernel $K_j$ of Eq.~(\ref{eq:kernel}) for this ghost point.}
\label{fig:points}
\end{figure}

In three dimensions, a ghost point has many fluid nodes near it, and each of them gives a ghost value through Eq.~(\ref{eq:oned}).
These values differ by the truncation errors of the extrapolations, but a ghost point can hold only one value.
We therefore take their average.
The candidates are the nodes of the $5 \times 5 \times 5$ stencil centered on the ghost point.
A candidate is used as a reference point if it satisfies three conditions (Fig.~\ref{fig:points}\pl{a}).
First, it must be a fluid node, $\psi > 0$.
Second, its distance from the ghost point must be smaller than a search radius $r$.
Third, the object point must be closer to the ghost point than to the candidate, i.e., $d_{1,i} < d_{2,i}$.
With the $N_\RP$ reference points $\RP_i$ $(i = 1, \dots, N_\RP)$, the ghost value is then
\begin{equation}
u_\GP = - \frac{1}{N_\RP} \sum_{i=1}^{N_\RP} \frac{d_{1,i}}{d_{2,i}}\, u_{\RP,i} . \label{eq:average}
\end{equation}
Throughout this paper, we use $r = 2.5 \Delta x$.

\begin{figure}[t]
\centering
\begin{tabular}{@{}c@{\hspace{0.05\textwidth}}c@{}}
\panel{a}{\includegraphics[width=0.37\textwidth]{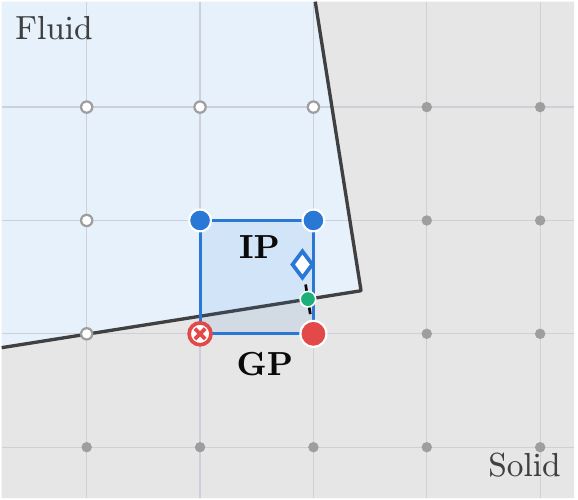}} &
\panel{b}{\includegraphics[width=0.37\textwidth]{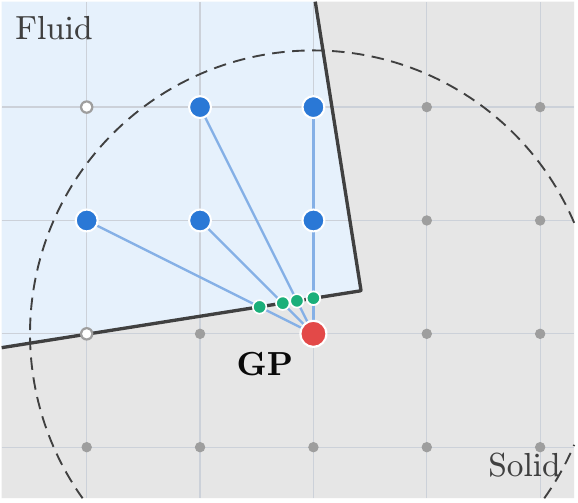}} \\[3pt]
\multicolumn{2}{c}{\includegraphics[width=0.80\textwidth]{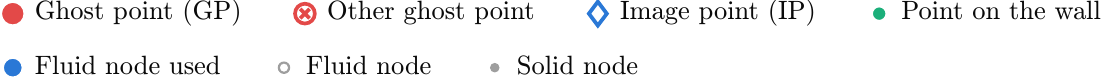}}
\end{tabular}
\caption{A ghost point (GP) at a concave corner, illustrated in two dimensions: \pl{a} conventional image-point reconstruction, in which the four nodes surrounding the image point (IP) include the ghost point itself and another ghost point (crossed circle); \pl{b} present method, in which all nearby fluid nodes satisfying the three conditions serve as reference points.}
\label{fig:corner}
\end{figure}

Equation~(\ref{eq:average}) has three useful properties.
First, it needs neither a wall normal nor an image point.
The method needs only the points where straight segments between grid nodes cross the wall, so that the ghost value is defined as soon as one reference point exists.
This also holds at sharp, thin or concave parts of the body, where a wall normal or an image-point stencil in the fluid may not be available (Fig.~\ref{fig:corner}).
Second, the ghost value is explicit.
Only fluid nodes appear on the right-hand side, so that a ghost value does not depend on other ghost values.
Consequently, no linear system or iteration is needed, and the result does not depend on the order in which the ghost points are processed.
Third, the weights are bounded.
Because $d_{1,i} < d_{2,i}$ in every term, the absolute values of the weights sum to less than one, and $|u_\GP| < \max_i |u_{\RP,i}|$.
Without this condition, the extrapolation~(\ref{eq:oned}) would be ill-conditioned for a reference point very close to the wall ($d_{2,i} \to 0$); the third condition excludes such points.

\subsection{Offline computation of the kernels}
\label{sec:kernel}

The quantities $d_{1,i}$ and $d_{2,i}$ and the selection of the reference points depend only on the geometry and the grid, not on the flow.
Equation~(\ref{eq:average}) can therefore be written as a weighted sum over the $5 \times 5 \times 5$ stencil $\mathcal{S}$ of the ghost point,
\begin{equation}
u_\GP = \sum_{j \in \mathcal{S}} K_j\, u_j , \qquad
K_j = \begin{cases} -\dfrac{d_{1,j}}{N_\RP\, d_{2,j}} & \text{if node $j$ is a reference point,}\\[6pt] 0 & \text{otherwise.} \end{cases}
\label{eq:kernel}
\end{equation}
The kernel $K$ has 125 weights and is computed before the simulation (Fig.~\ref{fig:points}b).
The kernels are generated in four steps.
First, the signed distance function $\psi$ is evaluated at the nodes of each velocity component.
For the sphere, it is known analytically.
For a body given as a triangulated surface (STL file), we compute the distance to the surface mesh and determine its sign with the generalized winding number \cite{jacobson2013,barill2018}.
Second, the ghost points are identified as the solid nodes with a fluid face neighbor.
Third, the object points are located.
For each ghost point, we consider the segments to the 124 other nodes of its stencil, and the wall crossing of each segment is located by bisection (ten iterations) on the tri-linear interpolant of $\psi$.
This gives $d_{1,j}$ and $d_{2,j}$.
Finally, the three conditions of Section~\ref{sec:average} are applied, and the weights are computed from Eq.~(\ref{eq:kernel}).
For each velocity component, the result is a list of the grid indices of the ghost points with 125 weights each.

Table~\ref{tab:kernel} lists the numbers of ghost points and reference points for the two geometries of this paper.
The number of ghost points is proportional to the surface area of the body measured in grid cells.
On average, a ghost point has 23--27 reference points.
The minimum was five for the sphere and one for the Stanford Bunny, so that every ghost point had at least one reference point.

\begin{table}[t]
\caption{Ghost points and reference points for $r = 2.5\Delta x$. The numbers of ghost points are given per velocity component (for the Stanford Bunny they differ slightly between the components). The Stanford Bunny is described in Section~\ref{sec:conv}.}
\label{tab:kernel}
\centering
\begin{tabular}{@{}lcccc@{}}
\toprule
 & & \multicolumn{3}{c}{$1/\Delta x$} \\
\cmidrule(l){3-5}
Geometry & Point attribute & 16 & 32 & 64 \\
\midrule
Sphere & ghost points & 604 & 2568 & 10484 \\
 & reference points per GP (min / mean / max) & 9 / 26.8 / 43 & 5 / 24.2 / 42 & 5 / 23.6 / 42 \\[2pt]
Stanford Bunny & ghost points & 953--965 & 4223--4244 & 17627--17660 \\
 & reference points per GP (min / mean / max) & 2 / 26.0 / 73 & 1 / 24.0 / 65 & 2 / 23.0 / 60 \\
\bottomrule
\end{tabular}
\end{table}

\subsection{Coupling with the flow solver}
\label{sec:coupling}

During the simulation, the ghost values are written directly into the velocity field by Eq.~(\ref{eq:kernel}).
This is done once at the beginning of each time step.
In addition, it is done again in each substep for the predicted velocity $\bm{u}^{\ast}$, after Eq.~(\ref{eq:pred}) and before the Poisson equation~(\ref{eq:poisson}).
A substep thus consists of four steps.
In step~1, the right-hand side of the predictor, i.e., all terms of Eq.~(\ref{eq:pred}) that are known from the previous substeps, is evaluated at all nodes and then set to zero at the ghost points.
In step~2, the implicit system of Eq.~(\ref{eq:pred}) is solved for $\bm{u}^{\ast}$ at all nodes.
In step~3, the ghost points of each velocity component are overwritten as follows:

\begin{equation}
u^{\ast}_\GP = \sum_j K_j u^{\ast}_j .  \label{eq:u_GP}
\end{equation}

Finally, in step~4, Eq.~(\ref{eq:poisson}) is solved, and the velocity and the pressure are corrected by Eq.~(\ref{eq:corr}) at all nodes.
In this procedure, the structure of the flow solver is unchanged.
The momentum and Poisson equations are solved on the whole Cartesian grid, including the interior of the body.
The same difference stencils are used everywhere, and the fast direct Poisson solver is retained.
The only additions, in steps~1 and~3, are loops over the list of ghost points.
Moreover, the update of step~3 is the same multiply--accumulate operation for every ghost point, and neither loop branches on the local geometry.

\subsection{Evaluation of the fluid force}
\label{sec:force}

The force on the body is obtained from the momentum balance of a control volume $V$ that encloses the body and whose surface $S$ lies entirely in the fluid:
\begin{equation}
\bm{F} = - \frac{\mathrm{d}}{\mathrm{d}t} \int_{V} \bm{u}\, \mathrm{d}V - \oint_{S} \left[(\bm{uu}) \cdot \bm{n} + p\, \bm{n} - \frac{1}{\Rey} \left( \bm{\nabla} \bm{u} + (\bm{\nabla} \bm{u})^{\mathrm{T}} \right) \cdot \bm{n} \right] \mathrm{d}S , \label{eq:force}
\end{equation}
where $\bm{n}$ is the outward normal of $S$.
The first term vanishes for a steady flow.
As the control volume, we use a cube of side 2 centered on the body, whose faces lie on cell faces of the grid.
All terms can then be evaluated at their staggered positions with second-order formulas.
In the present non-dimensionalization, the drag coefficient of the sphere is $C_D = 8 F_x / \pi$.

\section{Results and discussion}
\label{sec:results}

\subsection{Flow around a sphere: drag coefficient}
\label{sec:sphere}

\begin{figure}[t]
\centering
\begin{tabular}{@{}c@{\hspace{0.02\textwidth}}c@{}}
\panel{a}{\includegraphics[width=0.48\textwidth]{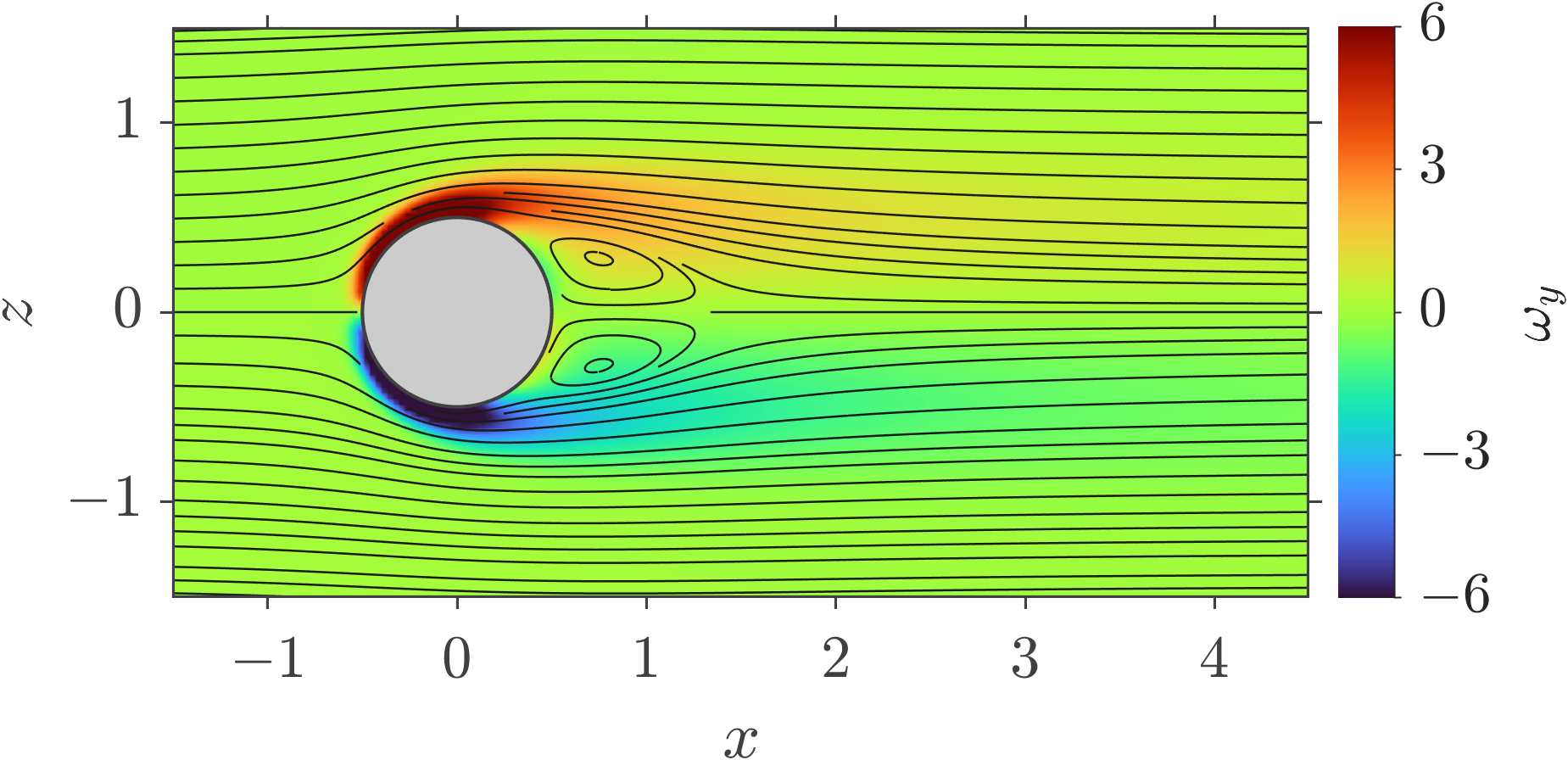}} &
\panel{b}{\includegraphics[width=0.48\textwidth]{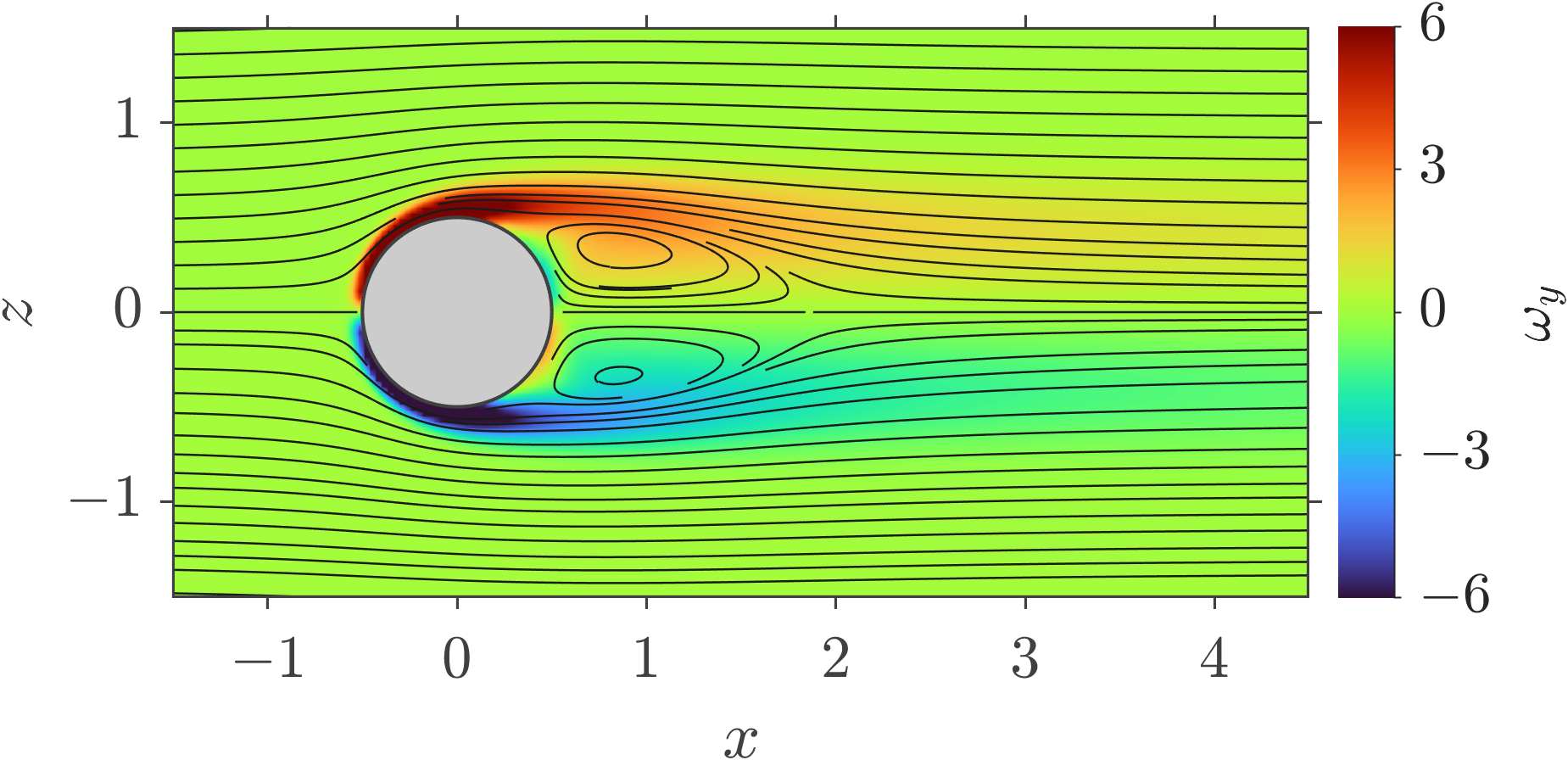}}
\end{tabular}
\caption{Steady flow around the sphere computed with $1/\Delta x = 32$ in the domain $L = 16$: vorticity component normal to the plane and streamlines in the center plane. \pl{a} $\Rey = 100$; \pl{b} $\Rey = 200$. The origin of the coordinates is shifted to the center of the sphere.}
\label{fig:wake}
\end{figure}

As the first test, we compute the steady, axisymmetric flow around a sphere at $\Rey = 25$, 50, 100, 150 and 200.
We restrict the comparison to this regime, because the steady flow loses its axial symmetry at $\Rey \approx 210$ \cite{johnson1999}.
The sphere has the diameter 1 and is placed at the center of a cubic domain of side $L$, where it is resolved with $1/\Delta x = 16$, 32 or 64 grid cells per diameter.
The grid is uniform in $x$ and $z$.
In $y$, it is uniform within a distance of 1 from the center of the sphere and stretched outside this region.
For all three resolutions, we use the domain sizes of $L = 8$ and 16.
The largest grid thus has $1024 \times 256 \times 1024$ cells ($1/\Delta x = 64$, $L = 16$).
The time step is $\Delta t = 0.02$ for $1/\Delta x = 16$ and 32, and 0.01 for $1/\Delta x = 64$.
Each computation starts from the uniform flow and is continued until $t = 120$.
The drag coefficient is evaluated from Eq.~(\ref{eq:force}).
At $t = 120$, the flow is steady: over the last five time units, the drag coefficient changes by less than $10^{-4}$~\%.
We therefore use its value at $t = 120$.
Fig.~\ref{fig:wake} shows the resulting steady flow at $\Rey = 100$ and 200.
The recirculation region behind the sphere is 0.84 and 1.34 diameters long, respectively.

\begin{table}[t]
\caption{Drag coefficient of the sphere obtained in each computation and its value extrapolated to an infinite domain ($L \to \infty$).}
\label{tab:dragruns}
\centering
\begin{tabular}{@{}ccccccc@{}}
\toprule
$1/\Delta x$ & Domain & $\Rey = 25$ & 50 & 100 & 150 & 200 \\
\midrule
16 & $8^3$ & 2.402 & 1.575 & 1.056 & 0.840 & 0.715 \\
 & $16^3$ & 2.355 & 1.549 & 1.042 & 0.830 & 0.706 \\
 & $L \to \infty$ & 2.339 & 1.540 & 1.037 & 0.826 & 0.704 \\[2pt]
32 & $8^3$ & 2.420 & 1.599 & 1.092 & 0.882 & 0.758 \\
 & $16^3$ & 2.372 & 1.573 & 1.076 & 0.870 & 0.749 \\
 & $L \to \infty$ & 2.356 & 1.564 & 1.071 & 0.866 & 0.746 \\[2pt]
64 & $8^3$ & 2.426 & 1.608 & 1.104 & 0.898 & 0.777 \\
 & $16^3$ & 2.377 & 1.581 & 1.088 & 0.886 & 0.767 \\
 & $L \to \infty$ & 2.361 & 1.572 & 1.083 & 0.881 & 0.764 \\
\bottomrule
\end{tabular}
\end{table}

Table~\ref{tab:dragruns} lists the results.
The drag coefficient decreases as the domain is enlarged.
The main reason is the blockage of the cross-section by the sphere, $(\pi/4) L^{-2}$, which is 1.2~\% for $L = 8$ and 0.3~\% for $L = 16$.
With $1/\Delta x = 32$, for example, the drag for $L = 8$ is higher than that for $L = 16$ by 1.2~\% at $\Rey = 200$ and by 2.0~\% at $\Rey = 25$.
We assume that the remaining error is proportional to $L^{-2}$, like the blockage.
The drag coefficient for an infinite domain then follows from these two domains,
\begin{equation}
C_{D,\infty} = C_{D,16} - \frac{C_{D,8} - C_{D,16}}{3} . \label{eq:cdinf}
\end{equation}
This extrapolation lowers the value for $L = 16$ by 0.4--0.7~\%.

The drag coefficient increases monotonically with the grid resolution at all Reynolds numbers.
The infinite-domain value rises by 0.7--6.0~\% from $1/\Delta x = 16$ to 32 and by a further 0.2--2.4~\% from 32 to 64.
This dependence on the resolution is stronger at the higher Reynolds numbers, where the boundary layer is thinner.

\begin{figure}[t]
\centering
\begin{tabular}{@{}c@{\hspace{0.03\textwidth}}c@{}}
\panel{a}{\includegraphics[width=0.47\textwidth]{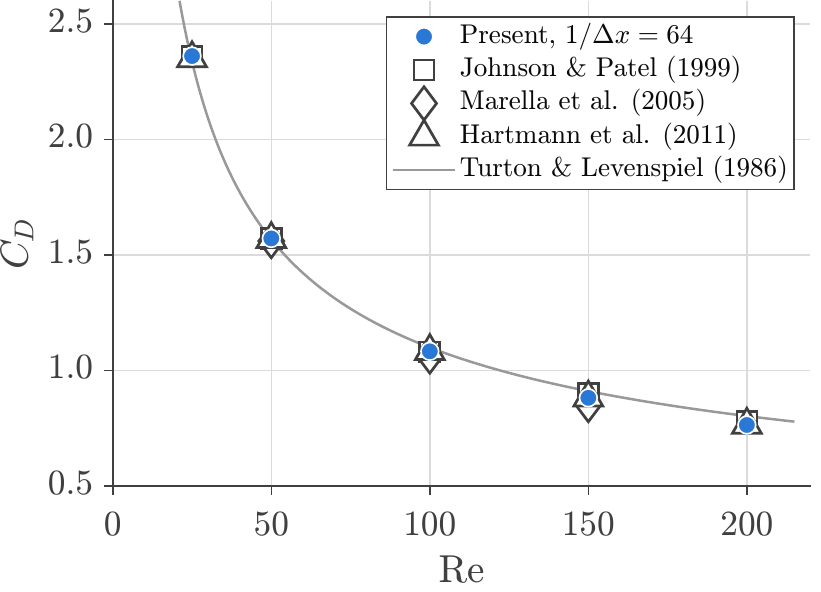}} &
\panel{b}{\includegraphics[width=0.47\textwidth]{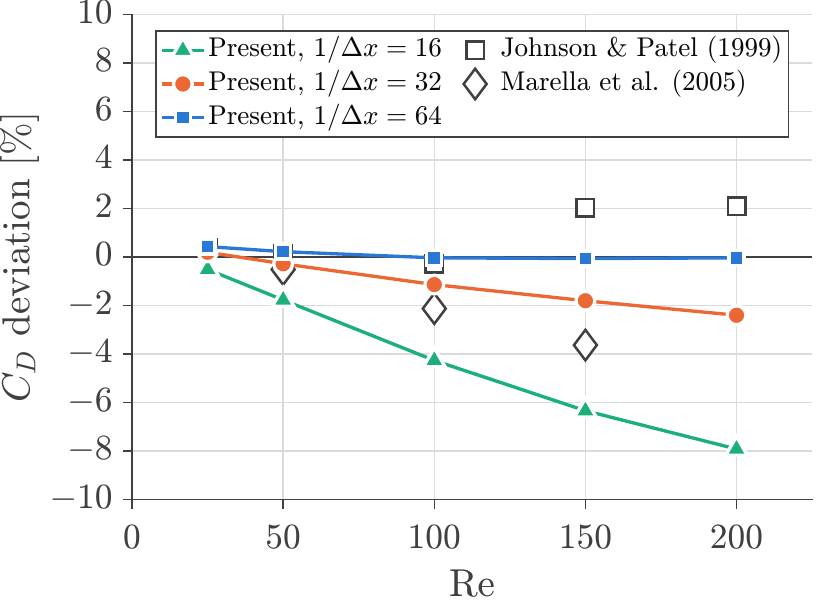}}
\end{tabular}
\caption{Drag coefficient of the sphere, extrapolated to an infinite domain: \pl{a} comparison of the present result for $1/\Delta x = 64$ with previous studies; \pl{b} deviation from the values of Hartmann et al.~\cite{hartmann2011} for the three grid resolutions, together with the other references.}
\label{fig:cd}
\end{figure}

Table~\ref{tab:dragcomp} and Fig.~\ref{fig:cd} compare the infinite-domain values with previous studies.
These are the computations of Johnson and Patel~\cite{johnson1999} on a body-fitted grid, of Marella et al.~\cite{marella2005} with a sharp-interface Cartesian grid method and of Hartmann et al.~\cite{hartmann2011} with a cut-cell method.
In addition, we show the empirical correlation of Turton and Levenspiel~\cite{turton1986}.
Relative to Hartmann et al.~\cite{hartmann2011}, the present drag coefficient deviates by $-2.4$~\% to $+0.2$~\% with $1/\Delta x = 32$ and by $-0.1$~\% to $+0.4$~\% with $1/\Delta x = 64$.
These deviations are no larger than the differences among the previous studies themselves.
For comparison, the values of Johnson and Patel~\cite{johnson1999} differ from those of Hartmann et al.~\cite{hartmann2011} by $-0.3$~\% to $+2.1$~\%, and those of Marella et al.~\cite{marella2005} by $-3.6$~\% to $-0.5$~\%.
With only 16 cells per diameter, however, the drag is underestimated by up to 7.9~\%.

\begin{table}[t]
\caption{Drag coefficient of the sphere, extrapolated to an infinite domain, compared with previous studies.}
\label{tab:dragcomp}
\centering
\begin{tabular}{@{}llccccc@{}}
\toprule
Study & Method & $\Rey = 25$ & 50 & 100 & 150 & 200 \\
\midrule
Johnson and Patel~\cite{johnson1999} & body-fitted grid & 2.36 & 1.57 & 1.08 & 0.90 & 0.78 \\
Marella et al.~\cite{marella2005} & sharp-interface method & -- & 1.56 & 1.06 & 0.85 & -- \\
Hartmann et al.~\cite{hartmann2011} & cut-cell method & 2.351 & 1.568 & 1.083 & 0.882 & 0.764 \\
Turton and Levenspiel~\cite{turton1986} & empirical correlation & 2.34 & 1.57 & 1.10 & 0.91 & 0.80 \\[2pt]
Present, $1/\Delta x = 16$ & ghost-cell IBM & 2.339 & 1.540 & 1.037 & 0.826 & 0.704 \\
Present, $1/\Delta x = 32$ & ghost-cell IBM & 2.356 & 1.564 & 1.071 & 0.866 & 0.746 \\
Present, $1/\Delta x = 64$ & ghost-cell IBM & 2.361 & 1.572 & 1.083 & 0.881 & 0.764 \\
\midrule
\multicolumn{2}{@{}l}{Deviation from Hartmann et al.~\cite{hartmann2011}, $1/\Delta x = 16$ [\%]} & $-0.5$ & $-1.8$ & $-4.3$ & $-6.3$ & $-7.9$ \\
\multicolumn{2}{@{}l}{Deviation from Hartmann et al.~\cite{hartmann2011}, $1/\Delta x = 32$ [\%]} & $+0.2$ & $-0.3$ & $-1.1$ & $-1.8$ & $-2.4$ \\
\multicolumn{2}{@{}l}{Deviation from Hartmann et al.~\cite{hartmann2011}, $1/\Delta x = 64$ [\%]} & $+0.4$ & $+0.2$ & $0.0$ & $-0.1$ & $0.0$ \\
\bottomrule
\end{tabular}
\end{table}

\subsection{Grid convergence of the flow field}
\label{sec:conv}

\begin{figure}[t]
\centering
\includegraphics[width=0.32\textwidth]{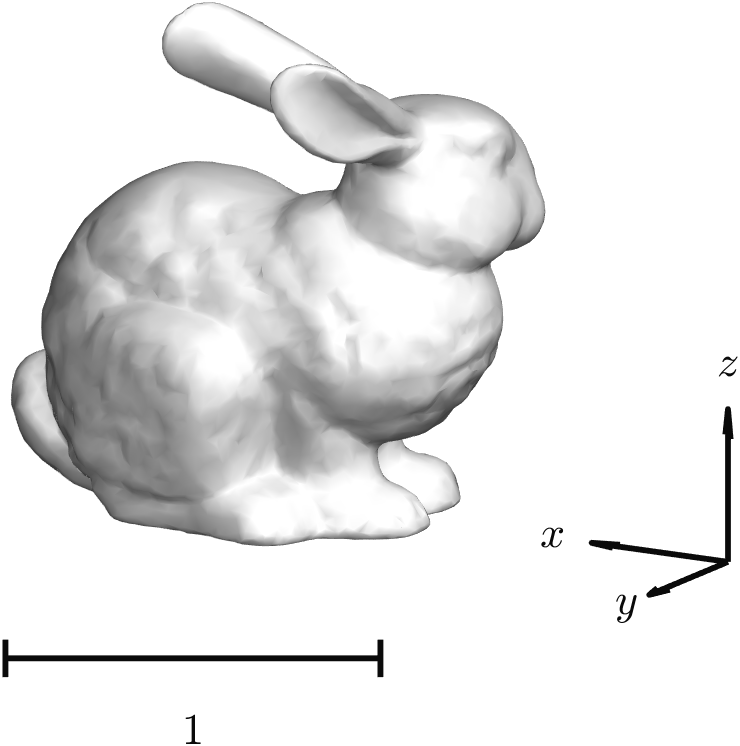}
\caption{The Stanford Bunny \cite{turk1994}, used in Sections~\ref{sec:conv} and~\ref{sec:bunny}, with the coordinate axes and a bar of the reference length.}
\label{fig:bunny}
\end{figure}

Next, we measure the order of accuracy by a grid refinement study for two bodies.
The first is the sphere.
The second is the Stanford Bunny \cite{turk1994}, a standard test geometry of computer graphics with concave regions and two thin, highly curved ears (Fig.~\ref{fig:bunny}).
Its surface mesh has 14,404 triangles, and we use it with a bounding box of $1.45 \times 1.11 \times 1.46$ (in $x$, $y$, $z$) in units of the reference length.
Each body is placed at the center of a cubic domain of side 4 with a uniform grid, $1/\Delta x = 8$, 16, 32 and 64.
The flow at $\Rey = 50$ is started impulsively from the uniform velocity.
We compare the solutions at $t = 1.2$, when the boundary layer and the separated region behind the body are still developing.
The reference solution is obtained on a grid with $1/\Delta x = 128$ ($512^3$ cells).
Since no exact solution is available, we assess the convergence by the difference from this reference solution, i.e., by self-convergence.
All grids use the same time step, $\Delta t = 10^{-3}$, so that the differences between the grids reflect the spatial discretization.
On each grid, the kernels are generated by the procedure of Section~\ref{sec:kernel}.

\begin{figure}[t]
\centering
\begin{tabular}{@{}c@{\hspace{0.03\textwidth}}c@{}}
\panel{a}{\includegraphics[width=0.45\textwidth]{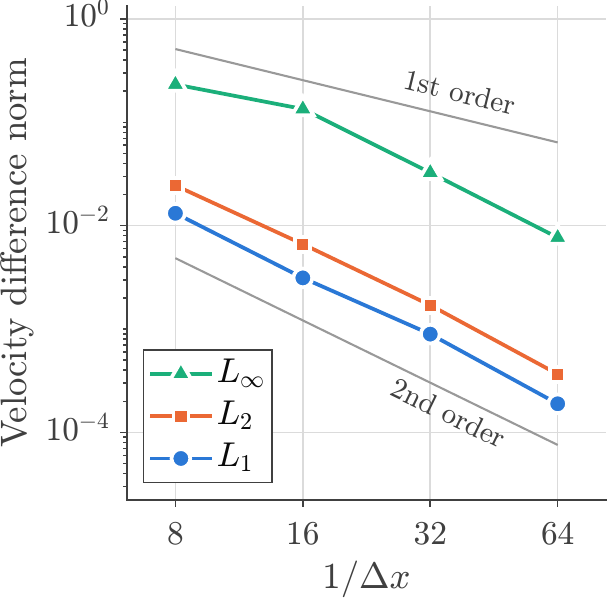}} &
\panel{b}{\includegraphics[width=0.45\textwidth]{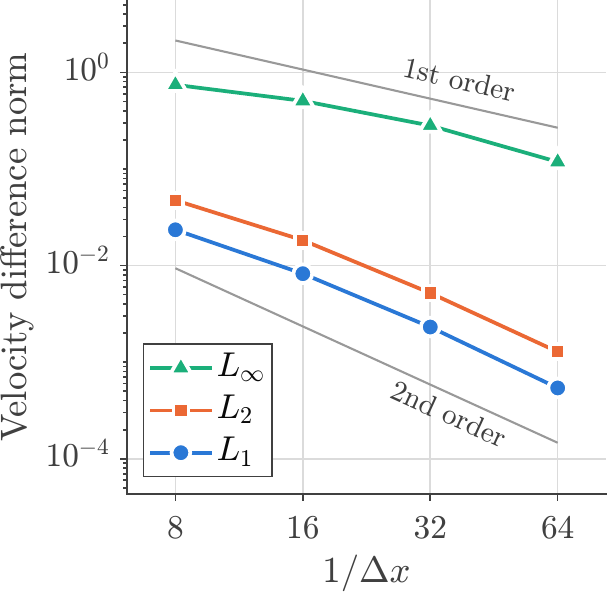}}
\end{tabular}
\caption{Difference of the velocity from the reference solution ($1/\Delta x = 128$) as a function of the grid resolution at $\Rey = 50$, $t = 1.2$: \pl{a} sphere; \pl{b} Stanford Bunny.}
\label{fig:conv}
\end{figure}

For each velocity component, the reference solution is interpolated tri-linearly to the nodes of the coarser grid.
We evaluate the difference at all fluid nodes of the three velocity components inside the box of side 2 centered on the body.
Nodes for which the interpolation would use a solid node of the reference grid are excluded.
Table~\ref{tab:conv} and Fig.~\ref{fig:conv} show the resulting mean absolute difference ($L_1$), root-mean-square difference ($L_2$) and maximum difference ($L_\infty$).

\begin{table}[t]
\caption{Difference of the velocity from the reference solution ($1/\Delta x = 128$) at $\Rey = 50$, $t = 1.2$, and the observed order of convergence between successive grids.}
\label{tab:conv}
\centering
\small
\setlength{\tabcolsep}{4.5pt}
\begin{tabular}{@{}llcccccccc@{}}
\toprule
 & & \multicolumn{4}{c}{Difference for $1/\Delta x =$} & & \multicolumn{3}{c}{Observed order} \\
\cmidrule(lr){3-6} \cmidrule(l){8-10}
 & Norm & 8 & 16 & 32 & 64 & & $8 \to 16$ & $16 \to 32$ & $32 \to 64$ \\
\midrule
Sphere & $L_1$ & $1.32 \times 10^{-2}$ & $3.13 \times 10^{-3}$ & $8.94 \times 10^{-4}$ & $1.89 \times 10^{-4}$ & & 2.08 & 1.81 & 2.24 \\
 & $L_2$ & $2.45 \times 10^{-2}$ & $6.63 \times 10^{-3}$ & $1.70 \times 10^{-3}$ & $3.67 \times 10^{-4}$ & & 1.88 & 1.96 & 2.21 \\
 & $L_\infty$ & $2.32 \times 10^{-1}$ & $1.34 \times 10^{-1}$ & $3.26 \times 10^{-2}$ & $7.67 \times 10^{-3}$ & & 0.79 & 2.04 & 2.09 \\[2pt]
Stanford Bunny & $L_1$ & $2.35 \times 10^{-2}$ & $8.24 \times 10^{-3}$ & $2.31 \times 10^{-3}$ & $5.41 \times 10^{-4}$ & & 1.51 & 1.83 & 2.09 \\
 & $L_2$ & $4.74 \times 10^{-2}$ & $1.83 \times 10^{-2}$ & $5.20 \times 10^{-3}$ & $1.28 \times 10^{-3}$ & & 1.37 & 1.82 & 2.02 \\
 & $L_\infty$ & $7.42 \times 10^{-1}$ & $5.06 \times 10^{-1}$ & $2.81 \times 10^{-1}$ & $1.18 \times 10^{-1}$ & & 0.55 & 0.85 & 1.25 \\
\bottomrule
\end{tabular}
\end{table}

For the sphere, the $L_1$ and $L_2$ norms of the difference decrease at second order over the whole range (observed orders of 1.8--2.2).
The maximum difference also decreases at second order once the sphere is resolved with 16 cells per diameter or more (orders of 2.0--2.1).
For the Stanford Bunny, the coarsest grid ($1/\Delta x = 8$) does not resolve the ears, so that the orders of the first refinement are lower.
Between $1/\Delta x = 16$ and 64, however, the $L_1$ and $L_2$ norms decrease at orders of 1.8--2.1.
Thus, since each one-dimensional extrapolation is second-order accurate, the averaging over many directions keeps second-order accuracy over most of the body, also for a body with thin and concave parts.

The maximum difference for the Stanford Bunny, in contrast, decreases only at orders of 0.6--1.2.
At the nodes more than $0.25$ away from the wall, on the other hand, it decreases at orders of 1.4--1.7.
Moreover, the maximum difference is found at the tip of an ear.
Such extremely thin or highly curved features make up only a very small part of the surface.
There, however, the accuracy of the present method drops, because an average of linear extrapolations cannot represent such a boundary accurately.
As a result, the maximum difference decreases more slowly than at second order, whereas the integral norms, which are set by the rest of the body, still decrease at second order.

In addition, we examine the pressure, for which no condition is imposed on the immersed boundary.
The pressure at the cell centers is compared with the reference solution, which is interpolated tri-linearly.
The pressure is set to zero mean over the fluid cells more than $2\Delta x$ away from the wall.
Table~\ref{tab:pressure} lists the $L_2$ norm of the difference for the first layer of fluid cells, whose centers lie within $\Delta x$ of the wall, for the cells more than $2\Delta x$ away from the wall, and for all fluid cells.
Between $1/\Delta x = 16$ and 64, the difference away from the wall decreases at orders of 1.5--2.2 for the sphere and 1.5--1.7 for the Stanford Bunny.
In the first layer, however, the orders are only 1.1--1.3 and 0.7--0.8, respectively.
Next to the wall, the convergence of the pressure thus drops to about first order.

\begin{table}[t]
\caption{$L_2$ norm of the difference of the pressure from the reference solution ($1/\Delta x = 128$) at $\Rey = 50$, $t = 1.2$, for the fluid cells at a distance $d$ from the wall, and the observed order of convergence between successive grids.}
\label{tab:pressure}
\centering
\small
\setlength{\tabcolsep}{4.5pt}
\begin{tabular}{@{}llcccccccc@{}}
\toprule
 & & \multicolumn{4}{c}{$L_2$ norm for $1/\Delta x =$} & & \multicolumn{3}{c}{Observed order} \\
\cmidrule(lr){3-6} \cmidrule(l){8-10}
 & Cells & 8 & 16 & 32 & 64 & & $8 \to 16$ & $16 \to 32$ & $32 \to 64$ \\
\midrule
Sphere & $0 < d \le \Delta x$ & $5.39 \times 10^{-2}$ & $1.73 \times 10^{-2}$ & $7.88 \times 10^{-3}$ & $3.25 \times 10^{-3}$ & & 1.64 & 1.13 & 1.28 \\
 & $d > 2\Delta x$ & $1.37 \times 10^{-2}$ & $3.59 \times 10^{-3}$ & $1.27 \times 10^{-3}$ & $2.81 \times 10^{-4}$ & & 1.93 & 1.50 & 2.18 \\
 & all fluid cells & $2.14 \times 10^{-2}$ & $4.90 \times 10^{-3}$ & $1.59 \times 10^{-3}$ & $3.68 \times 10^{-4}$ & & 2.13 & 1.62 & 2.12 \\[2pt]
Stanford Bunny & $0 < d \le \Delta x$ & $6.89 \times 10^{-2}$ & $4.54 \times 10^{-2}$ & $2.79 \times 10^{-2}$ & $1.64 \times 10^{-2}$ & & 0.60 & 0.70 & 0.77 \\
 & $d > 2\Delta x$ & $1.25 \times 10^{-2}$ & $6.76 \times 10^{-3}$ & $2.32 \times 10^{-3}$ & $7.22 \times 10^{-4}$ & & 0.89 & 1.55 & 1.68 \\
 & all fluid cells & $2.77 \times 10^{-2}$ & $1.27 \times 10^{-2}$ & $4.81 \times 10^{-3}$ & $1.70 \times 10^{-3}$ & & 1.12 & 1.40 & 1.50 \\
\bottomrule
\end{tabular}
\end{table}

% Fig. 8 is defined before the heading, so that it can be placed at the top of the page where this section
% begins; together with \FloatBarrier at the start of Section 3.4, all floats of this section then come before 3.4.
\begin{figure}[t]
\centering
\begin{tabular}{@{}c@{}}
\includegraphics[width=0.78\textwidth]{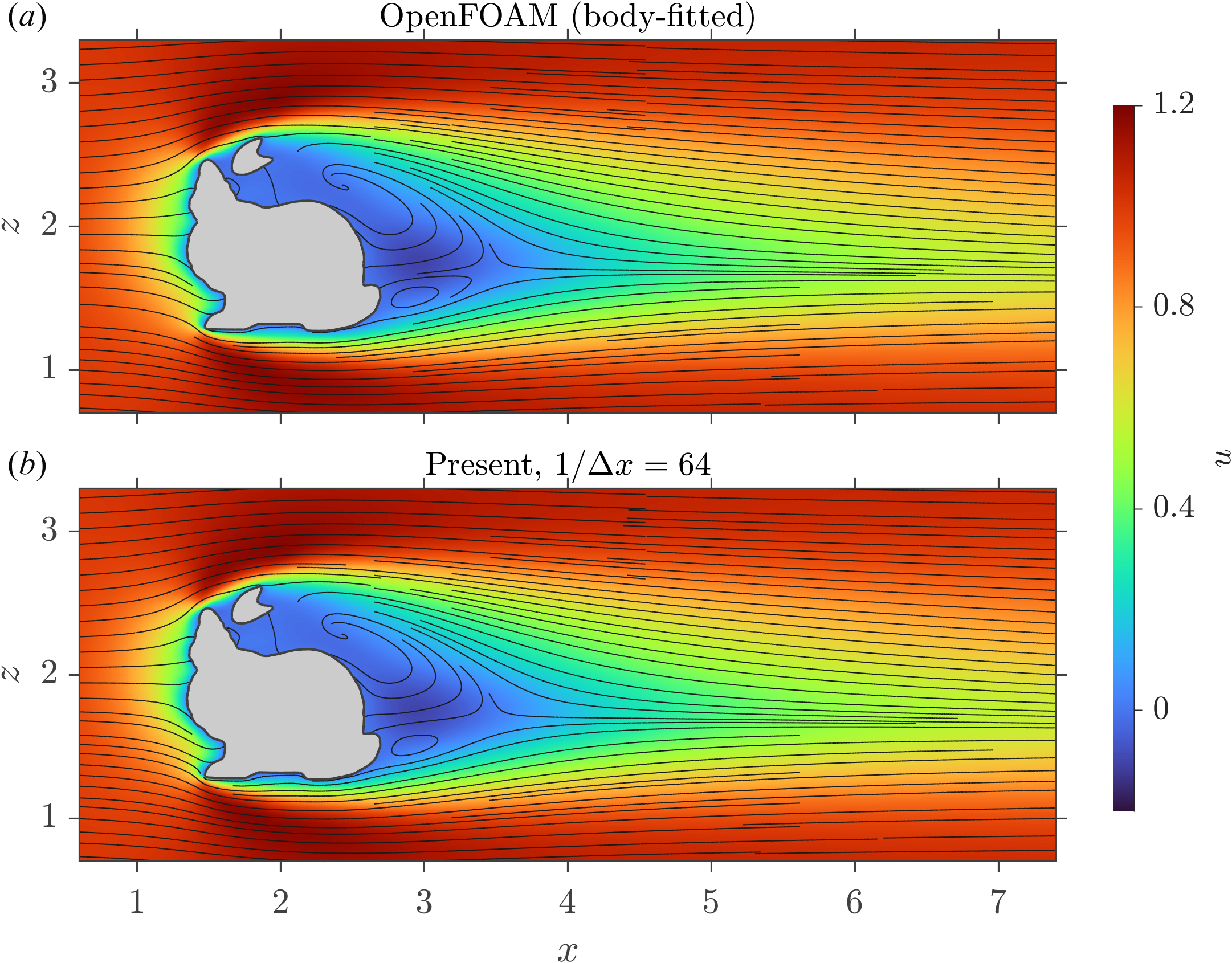}\\[4pt]
\includegraphics[width=0.98\textwidth]{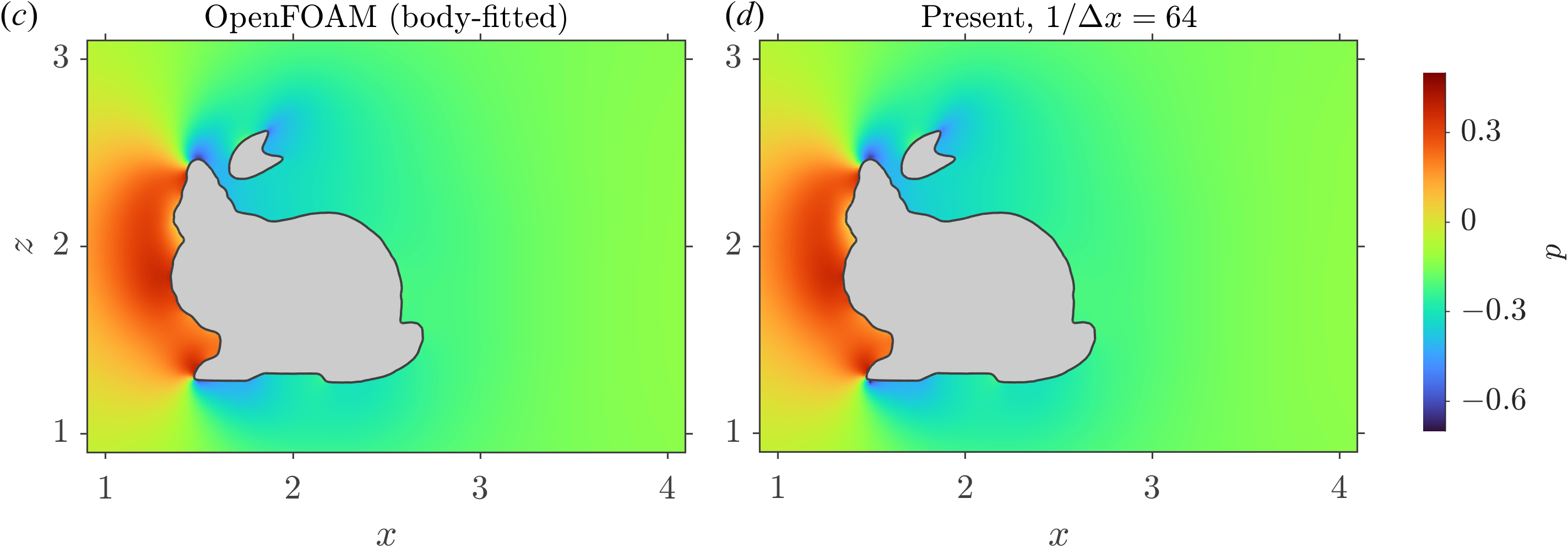}
\end{tabular}
\caption{Flow around the Stanford Bunny at $\Rey = 100$ in the center plane $y = 2$, computed with OpenFOAM on a body-fitted mesh, \pl{a} and \pl{c}, and with the present method ($1/\Delta x = 64$), \pl{b} and \pl{d}: \pl{a}, \pl{b} streamwise velocity and streamlines of the in-plane velocity; \pl{c}, \pl{d} pressure.}
\label{fig:bunnycmp}
\end{figure}

\subsection{Comparison of flow around Stanford Bunny with the reference field using a body-fitted mesh}
\label{sec:bunny}

Finally, we compare the present method with a simulation on a body-fitted mesh.
We compute the steady flow around the Stanford Bunny of Section~\ref{sec:conv} at $\Rey = 100$.
The domain is $8 \times 4 \times 4$, and the center of the bounding box of the body is located at $(2, 2, 2)$.
The flow enters with a uniform velocity at $x = 0$ and leaves the domain at $x = 8$.
The boundaries in $y$ are free-slip walls, and the boundaries in $z$ are periodic.
We use a uniform grid with $1/\Delta x = 64$, i.e., $512 \times 256 \times 256$ cells, and the time step 0.01.
The kernels are generated exactly as for the sphere, and no part of the surface is treated specially.
The computation is continued until $t = 120$, when the force no longer changes.
The force is evaluated from Eq.~(\ref{eq:force}).

The reference solution is computed with OpenFOAM~\cite{weller1998} (version 11) on a body-fitted mesh, in the same domain and with the same boundary conditions.
The mesh is generated by snappyHexMesh.
Its hexahedral cells have the size 1/8 far from the body.
They are refined to 1/32 around the body and its wake, to 1/64 near the body and to 1/128 at the surface, to which the mesh is snapped.
The mesh has $3.37 \times 10^6$ cells in total.
The steady equations are solved by the SIMPLEC algorithm, with a second-order upwind scheme for the convective term and central differences for the viscous term.
The iterations are continued until the residuals of the pressure and the velocity fall below $10^{-6}$ and $10^{-7}$, respectively.

\begin{table}[t]
\caption{Force coefficients $C_i = 8F_i/\pi$ of the Stanford Bunny at $\Rey = 100$. The values in parentheses are the deviations of the present result ($1/\Delta x = 64$) from the body-fitted solution.}
\label{tab:bunnyforce}
\centering
\begin{tabular}{@{}lccc@{}}
\toprule
 & $C_x$ & $C_y$ & $C_z$ \\
\midrule
OpenFOAM (body-fitted) & $1.8531$ & $-0.2987$ & $0.1219$ \\
Present, $1/\Delta x = 64$ & $1.8358$ ($-0.9$~\%) & $-0.2996$ ($+0.3$~\%) & $0.1213$ ($-0.5$~\%) \\
\bottomrule
\end{tabular}
\end{table}

Table~\ref{tab:bunnyforce} lists the force coefficients $C_i = 8F_i/\pi$ $(i = x, y, z)$, which use the same reference area as the drag coefficient of the sphere.
All three components agree with the body-fitted solution within 1~\%.

\begin{figure}[t]
\centering
\begin{tabular}{@{}c@{}}
\panel{a}{\includegraphics[width=0.78\textwidth]{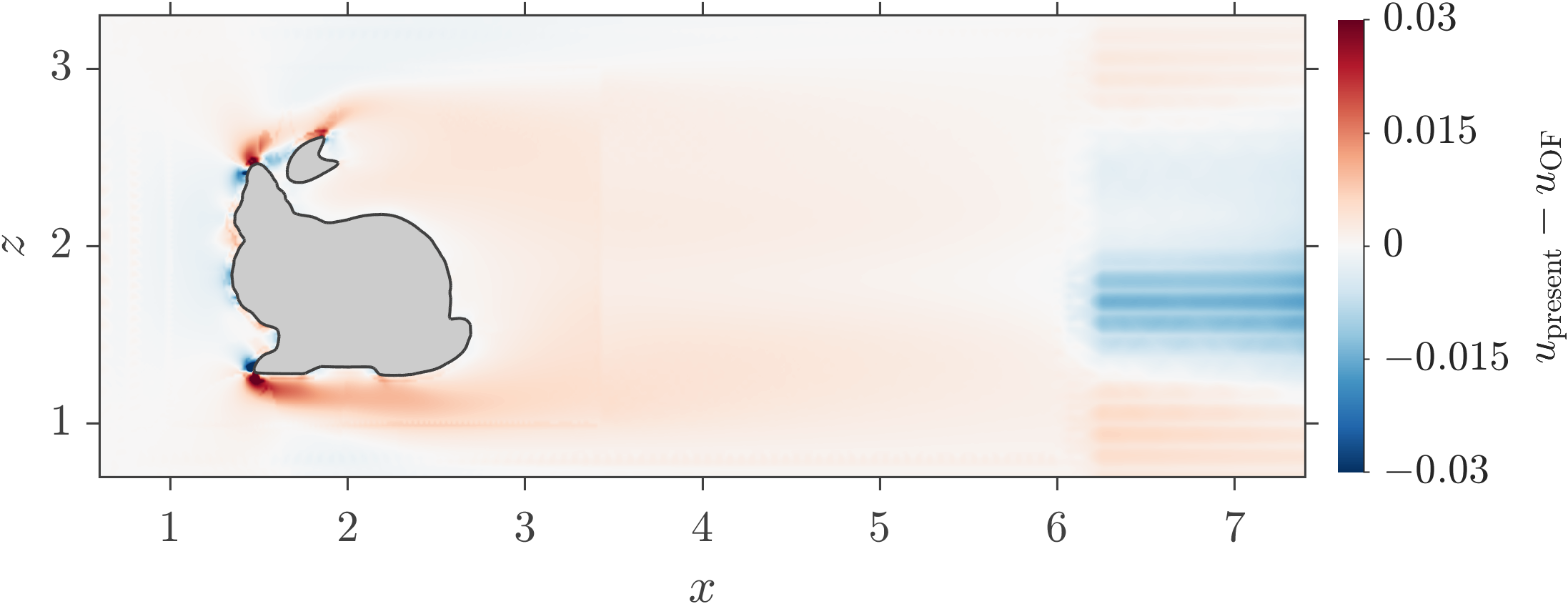}}\\[4pt]
\panel{b}{\includegraphics[width=0.56\textwidth]{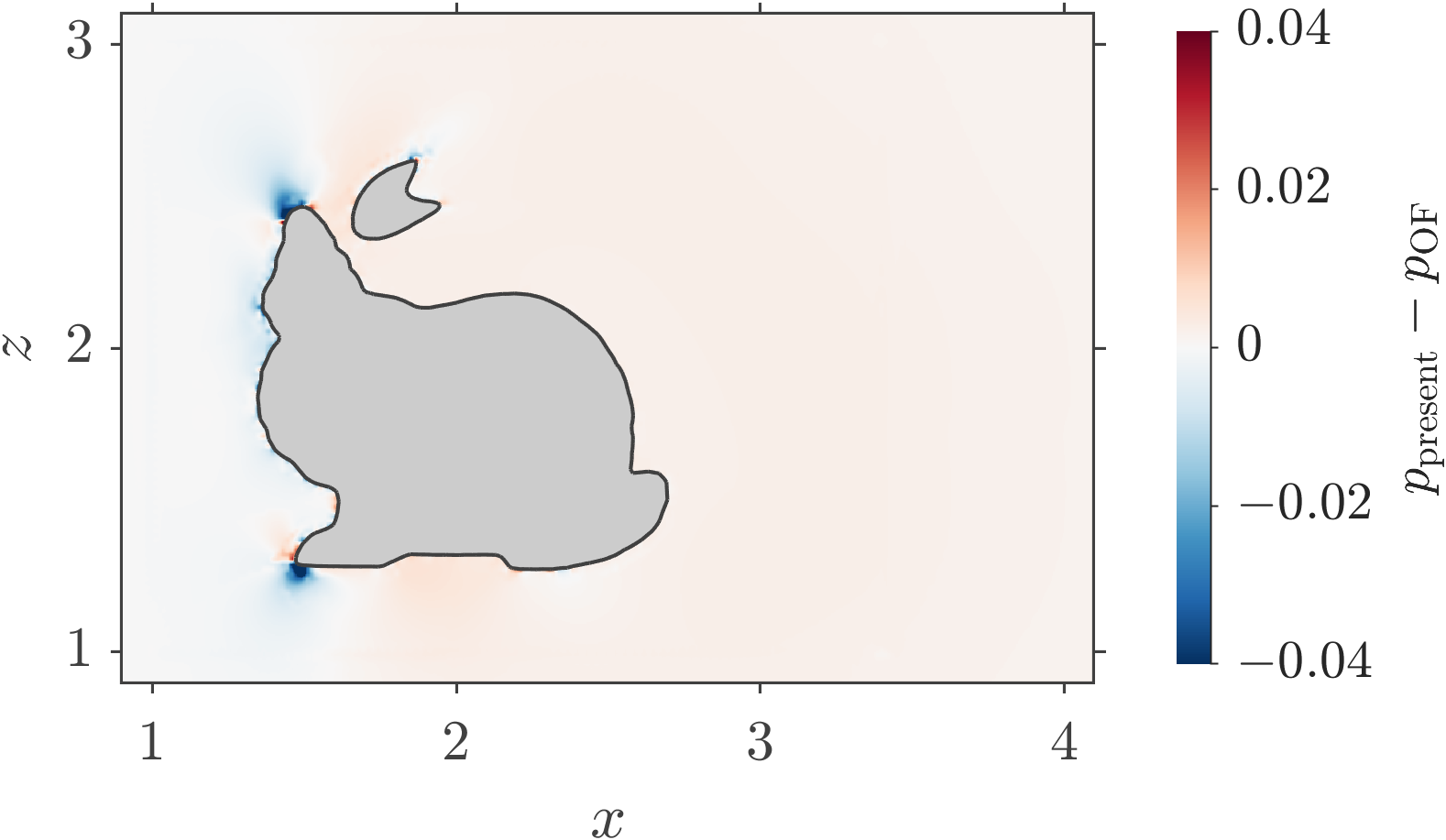}}
\end{tabular}
\caption{Difference between the present solution ($1/\Delta x = 64$) and the body-fitted solution in the center plane $y = 2$: \pl{a} streamwise velocity; \pl{b} pressure. $u_\mathrm{OF}$ and $p_\mathrm{OF}$ represent the streamwise velocity and pressure computed in OpenFOAM, respectively.}
\label{fig:bunnydif}
\end{figure}

Fig.~\ref{fig:bunnycmp} compares the streamwise velocity and the pressure in the center plane $y = 2$.
For this comparison, the present solution is averaged over the two cell layers next to the plane.
The pressure of both computations is set to zero mean upstream of the body.
The two solutions agree closely.
The recirculation region behind the body ends at $x = 3.45$ in the body-fitted solution and at $x = 3.44$ in the present one.
Moreover, the pressure is smooth up to the wall, although the present method imposes no condition for the pressure there.

Fig.~\ref{fig:bunnydif} shows the differences, which we evaluate at the cell centers of the present grid.
They are concentrated within a distance of about 0.1 from the wall, in particular at the upstream edges of the body.
There, the root-mean-square difference is 0.008 for $u$ and 0.008 for $p$.
Farther than 0.25 from the wall, it is only 0.003 and 0.002, respectively.
The steps at $x \approx 3.4$ and 6.2 in Fig.~\ref{fig:bunnydif}\pl{a} are the boundaries of the refinement regions of the body-fitted mesh.
Thus, also for a complex geometry with thin and concave parts, the present method reproduces the body-fitted solution closely.

\FloatBarrier
\subsection{Computational cost}
\label{sec:cost}

At run time, the boundary treatment is step~3 of Section~\ref{sec:coupling}.
For each ghost point of each velocity component, it is a multiply--accumulate operation with the weights of the kernel.
It is executed four times per time step: once at the beginning of the step and once in each of the three substeps.
In addition, the loop of step~1 adds one assignment per ghost point and substep.
The number of ghost points is small compared with the number of grid cells.
This is because it grows with the surface area of the body, i.e., as $\Delta x^{-2}$, whereas the number of cells grows as $\Delta x^{-3}$.
For example, the flow around the Stanford Bunny of Section~\ref{sec:bunny} has $3.4 \times 10^7$ cells and 52,924 ghost points in total for the three velocity components.
Similarly, the sphere with $1/\Delta x = 64$ in the domain $L = 8$ has $6.7 \times 10^7$ cells and 31,452 ghost points.
The kernels, on the other hand, are generated only once, before the simulation.
In addition to the signed distance function, this requires one bisection loop of ten iterations for each candidate node of a ghost point.
The loop locates the wall crossing on the segment between the two nodes.

Furthermore, the update at run time is the same operation for all ghost points, and the ghost points are independent of each other.
It is therefore well suited to vectorization and to parallel architectures, although the present implementation executes it serially.

\section{Conclusions}
\label{sec:concl}

We proposed a ghost-cell immersed boundary method based on multi-directional averaging.
In this method, the value at a ghost point is the average of one-dimensional linear extrapolations from all fluid nodes in the $5 \times 5 \times 5$ stencil around the ghost point that satisfy the conditions of Section~\ref{sec:average}, including a distance smaller than $2.5\Delta x$.
Each extrapolation passes through the point where the segment between the two nodes crosses the wall.
The construction therefore needs neither a wall normal nor an image point, and it is defined as soon as one such fluid node exists.
Moreover, it involves fluid nodes only, so that the ghost values are explicit and uncoupled.
Since the weights depend on the geometry alone, they are computed before the simulation from a signed distance function and stored as one kernel per ghost point.
As a result, the structure of the flow solver is unchanged.
Its only additions are two loops over the ghost points in each substep: one removes the ghost points from the predictor, and the other is the branch-free weighted sum.

The main findings are as follows.
First, we computed the steady flow around a sphere at $\Rey = 25$--200 and extrapolated the drag coefficient to an infinite domain.
The result deviates from that of Hartmann et al.~\cite{hartmann2011} by $-2.4$~\% to $+0.2$~\% with 32 cells per diameter and by $-0.1$~\% to $+0.4$~\% with 64 cells per diameter.
It increases monotonically with the grid resolution.
With only 16 cells per diameter, however, it is underestimated by up to 7.9~\%.
Second, in the self-convergence study, the velocity field converges at approximately second order in the $L_1$ and $L_2$ norms, both for a sphere and for the Stanford Bunny, with observed orders of 1.8--2.2 once the shape is resolved.
The maximum difference from the reference solution also decreases at second order for the sphere.
For the Stanford Bunny, in contrast, it decreases only at orders of 0.6--1.2, because it is set by a few nodes next to the wall.
Third, for the flow around the Stanford Bunny at $\Rey = 100$, the force coefficients agree with a body-fitted solution within 1~\% with $1/\Delta x = 64$.
The velocity and pressure fields also agree closely, although no condition for the pressure is imposed.
The kernels were generated exactly as for the sphere, and no part of the body, thin or concave, needed a specific treatment.
Finally, at run time, the boundary treatment is a multiply--accumulate operation per ghost point and velocity component, executed four times per time step.
Furthermore, the number of ghost points grows only with the surface area of the body.

On the other hand, the present formulation has limitations.
The convergence is not uniform: for a body with sharp, under-resolved features, the maximum difference next to the wall decreases much more slowly than the integral norms.
In addition, the pressure converges only at about first order next to the wall.
A subject of future work is a quantitative validation in unsteady flows, for example by the shedding frequency and the force fluctuations of the flow around a sphere.

\end{document}